\documentclass[prx, reprint, superscriptaddress]{revtex4-2}
\usepackage{amsmath}
\usepackage{amssymb}
\usepackage{graphicx}
\usepackage[caption=false, position=top, singlelinecheck=off, justification=raggedright]{subfig}
\usepackage{multirow}
\usepackage{color}
\usepackage{pst-node}
\usepackage[unicode]{hyperref}
\hypersetup{
	unicode=true,         	
	colorlinks=true,      	
	linkcolor=blue,		   	
	citecolor=blue,       	
	urlcolor=blue		   	
}

\begin{document}

\title{Terahertz amplifiers based on gain reflectivity in cuprate superconductors}

\author{Guido Homann}
\affiliation{Zentrum f\"ur Optische Quantentechnologien and Institut f\"ur Laserphysik, 
Universit\"at Hamburg, 22761 Hamburg, Germany}

\author{Jayson G. Cosme}
\affiliation{National Institute of Physics, University of the Philippines, Diliman, Quezon City 1101, Philippines}

\author{Ludwig Mathey}
\affiliation{Zentrum f\"ur Optische Quantentechnologien and Institut f\"ur Laserphysik, 
Universit\"at Hamburg, 22761 Hamburg, Germany}
\affiliation{The Hamburg Centre for Ultrafast Imaging, Luruper Chaussee 149, 22761 Hamburg, Germany}

\date{\today}
\begin{abstract}
We demonstrate that parametric driving of suitable collective modes in cuprate superconductors results in a reflectivity $R>1$ for frequencies in the low terahertz regime. We propose to exploit this effect for the amplification of coherent terahertz radiation in a laser-like fashion. As an example, we consider the optical driving of Josephson plasma oscillations in a monolayer cuprate at a frequency that is blue-detuned from the Higgs frequency. Analogously, terahertz radiation can be amplified in a bilayer cuprate by driving a phonon resonance at a frequency slightly higher than the upper Josephson plasma frequency. We show this by simulating a driven-dissipative $U(1)$ lattice gauge theory on a three-dimensional lattice, encoding a bilayer structure in the model parameters. We find a parametric amplification of terahertz radiation at zero and nonzero temperature.
\end{abstract}
\maketitle

\section{Introduction}
\begin{figure}[!b]
	\includegraphics[scale=1]{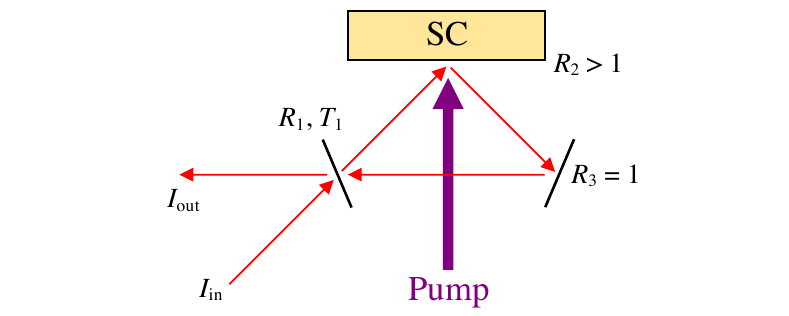}
	\caption{Setup of an optical parametric oscillator using a superconductor (SC) with reflectivity $R_2 > 1$ as a gain medium.}
	\label{fig:fig1} 
\end{figure}

Coherent radiation sources in the terahertz regime have applications in spectroscopy and imaging in numerous fields, such as biology and medical diagnostics, nondestructive evaluation, and solid state research \cite{Ferguson2002, Tonouchi2007, Pickwell2008, Hwang2015, Mittleman2018}. While significant progress has been made in the development of powerful terahertz sources \cite{Williams2007, Kumar2011, Asada2008, Feiginov2014, Ozyuzer2007, Welp2013, Kakeya2016, Kleiner2019, Cattaneo2021}, further development of terahertz technologies is imperative to close the `terahertz gap', particularly in the range between 0.5 and 1.5~THz \cite{Welp2013, Kakeya2016}.
In this work, we propose the design of an optical parametric oscillator \cite{Giordmaine1965, Akhmanov1965, Duarte2016} in the low-terahertz regime, i.e., $\sim$1~THz, to be utilized as an optical amplifier in a laser-like operation. We base this design on a general strategy to control the reflectivity of solids. The central mechanism is to use a collective mode with a nonlinear coupling to the electromagnetic field for parametric amplification. We apply this mechanism to cuprate superconductors and propose a laser-like setup for the amplification of terahertz radiation. In this setup, a light-driven superconductor with reflectivity $R_2= R(\omega_{\mathrm{pr}})>1$ serves as one of three mirrors forming an optical resonator as depicted in Fig.~\ref{fig:fig1}. The probe with frequency $\omega_{\mathrm{pr}}$ enters the resonator through a partially transparent mirror with reflectivity $R_1 < 1$ and transmissivity $T_1>0$. The third mirror is assumed to have perfect reflectivity $R_3= 1$. For $R_1 R_2 < 1$, the intensity ratio of the outgoing and the ingoing signal is given by
\begin{equation}
\frac{I_{\mathrm{out}}}{I_{\mathrm{in}}}= \frac{R_2 T_1^2}{1- R_1 R_2} + R_1 .
\end{equation}
The gain condition of this setup, $R_2 > 1/R_1$, is reflected by the divergence of $I_{\mathrm{out}}$ for $R_1 R_2 \rightarrow 1$. Above this threshold, the gain saturates once the probe signal enters the nonlinear response regime such that $R(\omega_{\mathrm{pr}})$ decreases. 

As our proposal requires pump lasers with field strengths of several hundred kV/cm, we suggest the following strategy towards its technical realization. The first step would be to test the proposed amplification mechanism using a pump pulse with a duration of $\sim$1~ps. Following the interpretation of the measurements in Ref.~\cite{Buzzi2021}, we propose to achieve better overlap of the pump and probe laser fields in the material by varying the incident angle of the probe pulse. We note that observing a net reflectivity gain from a light-driven superconductor would also be interesting from a purely scientific point of view. The next step would be a terahertz amplifier that is operated in pulsed fashion. Extending the duration of the pump pulse to $\sim$1~ns, as in Ref.~\cite{Budden2021}, would allow for several roundtrips of the probe pulse in an optical resonator with a pathlength of a few centimeters.

In this paper, we first demonstrate parametric amplification of terahertz radiation in monolayer cuprates using a gauge-invariant two-mode model with a cubic coupling process of the Higgs and plasma modes \cite{Homann2020, Homann2021}. We find that driving plasmonic excitations blue-detuned from the Higgs mode leads to a reflectivity $R>1$ for probe frequencies below the Josephson plasma edge. As a second example, we consider a periodic modulation of the interlayer coupling in bilayer cuprates, which models a periodic excitation of a phonon mode. Here, the low-frequency reflectivity is larger than 1 when the frequency of the excited phonon mode is blue-detuned from the upper Josephson plasma frequency. We implement a three-dimensional $U(1)$ lattice gauge theory with anisotropic lattice parameters to simulate this scenario at nonzero temperature. Our calculations show that phonon mediated amplification of terahertz radiation is effective at temperatures up to $\sim$20\% of the critical temperature $T_c$. Optical amplification in light-driven solids was also discussed in Refs.~\cite{Chiriaco2018, Cartella2018, Sugiura2019, Buzzi2021, Broers2021, Michael2021}.

The key requirement for the amplification mechanism presented in this work is a cubic coupling term of the form $\phi \theta^2$ in the Lagrangian, where $\theta$ is the plasma mode and $\phi$ represents another collective mode. Note that the plasma mode directly couples to the electric field $E$. When one applies pump and probe processes to the system, as sketched in Fig.~\ref{fig:fig2}(a), there are two scenarios for a parametric amplification of the probe. In one scenario, the pump directly excites the plasma mode $\theta$ at a frequency that is blue-detuned from the eigenfrequency of the mode $\phi$, which acts as an idler mode. Alternatively, the pump primarily couples to the collective mode $\phi$. An enhanced response is then achieved for a pump frequency $\omega_{\mathrm{dr}}$ that is blue-detuned with respect to the plasma frequency. Here, the plasma mode serves as the idler mode. In both cases, the probe couples to the plasma mode, and its frequency should be $\omega_{\mathrm{pr}}= \omega_{\mathrm{dr}} - \omega_{\mathrm{r}}$, where $\omega_{\mathrm{r}}$ denotes the eigenfrequency of the idler mode. Thus, three-wave mixing of the probe with the pump and resonant excitations of the idler mode induces the amplification of the probe signal. An important signature of this effect is a negative peak in the real part of the optical conductivity $\sigma_1$ at $\omega_{\mathrm{pr}}= \omega_{\mathrm{dr}} - \omega_{\mathrm{r}}$.

\section{Higgs mode mediated amplification in light-driven monolayer cuprates}
Josephson plasma oscillations are characteristic excitations of cuprate superconductors \cite{Koyama1996, Marel2001, Koyama2002, Dulic2001, Shibata1998}, corresponding to the tunneling of Cooper pairs between copper-oxide layers. The Higgs mode, on the other hand, describes amplitude oscillations of the superconducting order parameter \cite{Matsunaga2014, Tsuji2015, Katsumi2018, Chu2020, Shimano2020, Schwarz2020, Seibold2021}. While plasma modes directly couple to the electromagnetic vector potential, the Higgs mode has no linear coupling to electromagnetic fields in a system with approximate particle-hole symmetry \cite{Varma2002, Pekker2015}.
A two-mode model of a light-driven monolayer cuprate at zero temperature was derived in Refs.~\cite{Homann2020, Homann2021}. The underlying Lagrangian includes a cubic term $\sim$$h \theta^2$, coupling the plasma mode $\theta$ and the Higgs mode $h$. The equations of motion read
\begin{align}
	\ddot{\theta} + \gamma_{\mathrm{J}} \dot{\theta} + \omega_{\mathrm{J}}^2 \sin(\theta)(1+h)^2 &= j , \label{eq:eom1} \\
	\begin{split}
	\ddot{h} + \gamma_{\mathrm{H}} \dot{h} + \omega_{\mathrm{H}}^2 \left( h + \frac{3}{2}h^2 + \frac{1}{2} h^3 \right) \\
	+ 2\alpha \omega_{\mathrm{J}}^2\left[1-\cos(\theta) \right](1+h) &= 0 , \label{eq:eom2}
	\end{split}
\end{align}
where $\omega_{\mathrm{H}}$ is the Higgs frequency, $\omega_{\mathrm{J}}$ is the plasma frequency, and $\gamma_{\mathrm{H}}$ and $\gamma_{\mathrm{J}}$ are damping coefficients. The capacitive coupling constant $\alpha$ is of the order of 1 in cuprate superconductors \cite{Machida2004}. The interlayer current $j(t)= j_{\mathrm{dr}}(t) + j_{\mathrm{pr}}(t)$ is induced by an external electric field polarized along the $c$~axis of the crystal and describes the pump and probe processes. A monochromatic pump with field strength $E_0$ gives rise to $j_{\mathrm{dr}}(t)= (-2e d \omega_{\mathrm{dr}} E_0/\hbar \epsilon_{\infty}) \sin(\omega_{\mathrm{dr}} t)$, where $-2e$ is the Cooper pair charge, $d$ is the interlayer spacing, and $\epsilon_{\infty}$ is the background dielectric constant of the material. To calculate the optical conductivity, we include a weak probe current $j_{\mathrm{pr}}(t)$ and evaluate the Fourier components $j(\omega_{\mathrm{pr}})$ and $\theta(\omega_{\mathrm{pr}})$ in the steady state. The conductivity is given by $\sigma(\omega_{\mathrm{pr}})= i \epsilon_{\infty} \epsilon_0 j(\omega_{\mathrm{pr}})/\omega_{\mathrm{pr}} \theta(\omega_{\mathrm{pr}})$, as follows from the Josephson relation $\dot{\theta}= 2edE/\hbar$ \cite{Josephson1962}.

\begin{figure}[!b]
	\includegraphics[scale=1]{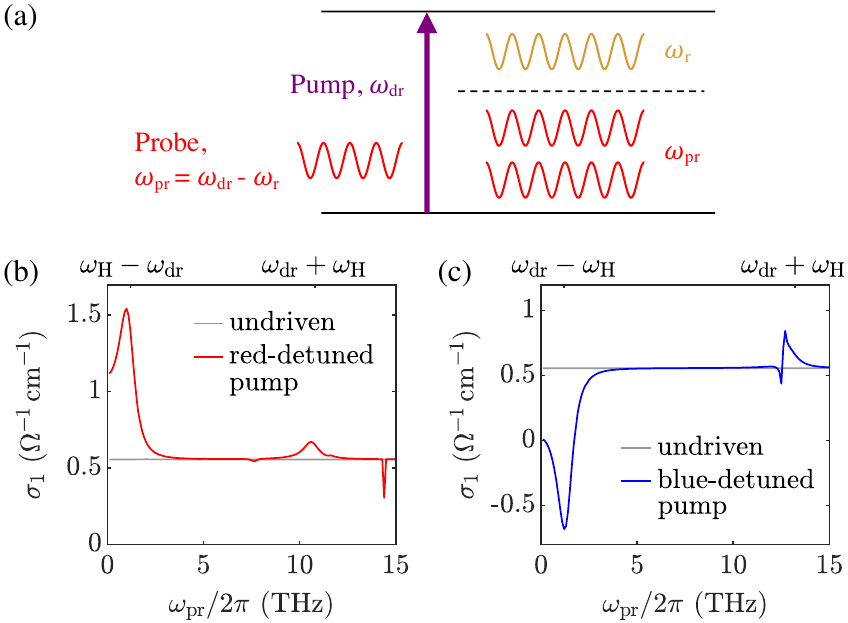}
	\caption{Parametric amplification of terahertz radiation in a solid with a plasma mode $\theta$ and a collective mode $\phi$ that are nonlinearly coupled. (a) Schematic illustration of the amplification process. The pump laser excites the plasma mode $\theta$ (or the collective mode $\phi$) with frequency $\omega_{\mathrm{dr}}$. The pump signal is down-converted to the lower frequency $\omega_{\mathrm{pr}}= \omega_{\mathrm{dr}}-\omega_{\mathrm{r}}$ of the probe by simultaneous excitation of the collective mode $\phi$ (or the plasma mode $\theta$) at its eigenfrequency $\omega_{\mathrm{r}}$. The numerical results in (b) and (c) are obtained for a monolayer cuprate, in which Josephson plasma oscillations are driven by the pump laser and the Higgs mode is the idler mode. (b) When the pump frequency is red-detuned from the Higgs frequency $\omega_{\mathrm{r}} \equiv \omega_{\mathrm{H}}$, a probe with $\omega_{\mathrm{pr}} \approx \omega_{\mathrm{H}}-\omega_{\mathrm{dr}}$ is attenuated as indicated by the positive peak in the real part $\sigma_1$ of the optical conductivity. The pump frequency is $\omega_{\mathrm{dr}}/2\pi= 4.8~\mathrm{THz}$ and the pump strength is $E_0= 150~\mathrm{kV/cm}$. (c) A blue-detuned pump frequency leads to an amplification of the probe for $\omega_{\mathrm{pr}} \approx \omega_{\mathrm{dr}}-\omega_{\mathrm{H}}$, corresponding to a negative peak in $\sigma_1$. In this case, the pump frequency is $\omega_{\mathrm{dr}}/2\pi= 7.2~\mathrm{THz}$ and the pump strength is $E_0= 300~\mathrm{kV/cm}$. The probe strength is $E_{\mathrm{pr}}= 1~\mathrm{kV/cm}$ in both cases. The Josephson plasma frequency is $\omega_{\mathrm{J}}/2\pi= 2~\mathrm{THz}$ and the Higgs frequency is $\omega_{\mathrm{H}}/2\pi= 6~\mathrm{THz}$. The remaining parameters are $\gamma_{\mathrm{J}}/2\pi= 0.25~\mathrm{THz}$, $\gamma_{\mathrm{H}}/2\pi= 1~\mathrm{THz}$, $\alpha=1$, $\epsilon_{\infty}=4$, and $d= 10~\text{\AA}$.}
	\label{fig:fig2} 
\end{figure}

In Figs.~\ref{fig:fig2}(b) and \ref{fig:fig2}(c), we present numerical results for the real part of the optical conductivity of a monolayer cuprate with Josephson plasma frequency $\omega_{\mathrm{J}}/2\pi= 2~\mathrm{THz}$ and Higgs frequency $\omega_{\mathrm{H}}/2\pi= 6~\mathrm{THz}$. For a pump frequency that is red-detuned with respect to the Higgs frequency, $\sigma_1$ exhibits a pronounced absolute maximum at $\omega_{\mathrm{pr}} \approx \omega_{\mathrm{H}}-\omega_{\mathrm{dr}}$ and a local maximum at $\omega_{\mathrm{pr}} \approx \omega_{\mathrm{H}} + \omega_{\mathrm{dr}}$. The peak at $\omega_{\mathrm{pr}} \approx \omega_{\mathrm{H}}-\omega_{\mathrm{dr}}$ corresponds to an excitation of the Higgs mode via resonant two-photon processes, whereas a probe with $\omega_{\mathrm{pr}} \approx \omega_{\mathrm{H}} + \omega_{\mathrm{dr}}$ amplifies the pump signal and simultaneously excites the Higgs mode. The minimum slightly below $15~\mathrm{THz}$ results from the coupling of the probe to the third-harmonic of the pump.

For a blue-detuned pump frequency, we find $\sigma_1<0$ for low probe frequencies. The mininum at $\omega_{\mathrm{pr}} \approx \omega_{\mathrm{dr}} - \omega_{\mathrm{H}}$ indicates a resonant amplification of the probe due to a down-conversion of the pump by simultaneous excitation of the Higgs mode. The conductivity displays a maximum at $\omega_{\mathrm{pr}} \approx \omega_{\mathrm{dr}} + \omega_{\mathrm{H}}$, similarly to the case of a red-detuned pump frequency, while the third-harmonic of the pump is outside the plotted frequency range. In Appendix~\ref{sec:analytical}, we provide an analytical estimate of $\sigma(\omega_{\mathrm{pr}}= \omega_{\mathrm{dr}}-\omega_{\mathrm{H}})$ based on a perturbative expansion for weak pump-probe strengths. Our analytical estimate is in qualitative agreement with the numerical results.

In the following, we focus on pump frequencies that are blue-detuned from the Higgs frequency. As we shall see below, a negative conductivity $\sigma_1$ implies a reflectivity $R>1$ at low frequencies. The reflectivity at normal incidence is obtained from the optical conductivity via the Fresnel equation
\begin{equation} \label{eq:reflectivity}
R(\omega)= \left| \frac{1- n(\omega)}{1+ n(\omega)} \right|^2 .
\end{equation}
The refractive index $n(\omega)= \pm \sqrt{\epsilon(\omega)}$ is a function of the dielectric permittivity $\epsilon(\omega)= \epsilon_{\infty} + i \sigma(\omega)/\epsilon_0 \omega$. The sign of the refractive index for a given frequency is fixed by causality \cite{Skaar2006, Nistad2008}. We choose the positive sign unless both the real part and the imaginary part of $\epsilon(\omega)$ are negative. Thus, the electric field penetrates the bulk for frequencies above the plasma edge around $\omega_{\mathrm{J}}/\sqrt{2}$, while it is screened for lower frequencies. This is the characteristic response of a Josephson plasma, also in the presence of a periodic drive \cite{Michael2021}.

\begin{figure}[!t]
	\includegraphics[scale=1]{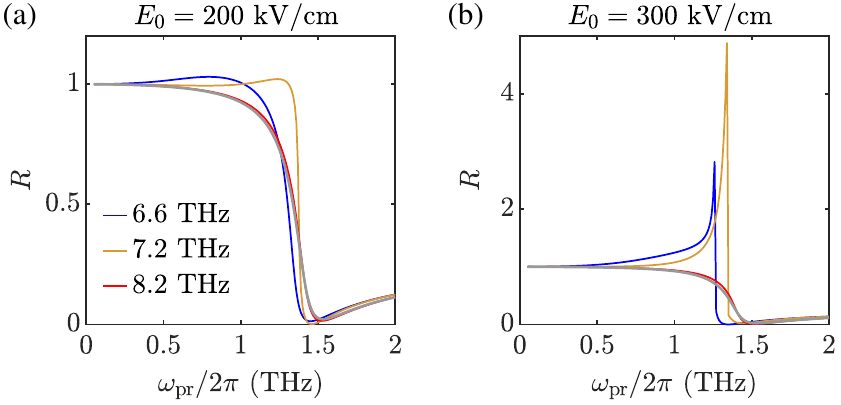}
	\caption{Higgs mode mediated amplification of terahertz radiation in a monolayer cuprate. The reflectivity at normal incidence is shown for two choices of the pump strength: (a) $E_0= 200~\mathrm{kV/cm}$, (b) $E_0= 300~\mathrm{kV/cm}$. The pump frequencies for the higher pump strength in (b) are the same as indicated for the lower pump strength in (a). Gray lines correspond to the undriven case. The probe strength is $E_{\mathrm{pr}}= 1~\mathrm{kV/cm}$. The model parameters are the same as in Fig.~\ref{fig:fig2}.}
	\label{fig:fig3} 
\end{figure}

Figure~\ref{fig:fig3} displays the low-frequency reflectivity for different strengths and frequencies of the optical pump applied to the same monolayer cuprate as before. For pump frequencies that are slightly blue-detuned from the Higgs frequency, the reflectivity is larger than 1 at probe frequencies below the plasma edge. This is an immediate consequence of the negative $\sigma_1$ at low probe frequencies in those cases. As expected, the enhancement of the reflectivity is more pronounced for the stronger pump in Fig.~\ref{fig:fig3}(b) than for the weaker pump in Fig.~\ref{fig:fig3}(a). The amplification mechanism is particularly effective if the detuning $\omega_{\mathrm{dr}}-\omega_{\mathrm{H}}$, and thus the minimum of $\sigma_1$, approaches the plasma edge frequency, as is the case for $\omega_{\mathrm{dr}}/2\pi= 7.2~\mathrm{THz}$. We note that the plasma edge is shifted to a slightly lower frequency by the pump, corresponding to a small reduction of the time-averaged superconducting order parameter. The magnitude of this shift increases with increasing pump strength and decreasing detuning of the pump frequency from the Higgs frequency.

\section{Phonon mediated amplification in bilayer cuprates}
We now turn to our second example of parametric amplification of terahertz radiation in cuprate superconductors. While the Higgs mode is strongly damped in the cuprates in general \cite{Katsumi2018, Chu2020, Peronaci2015}, phononic excitations have picosecond lifetimes, such as vibrations of apical oxygen atoms in YBa$_2$C$_3$O$_{7-\delta}$ (YBCO) \cite{Mankowsky2014, Mankowsky2015}. In the following, we consider the scenario in which the pump laser periodically modulates the Josephson coupling between the copper-oxide layers as it resonantly couples to a phonon mode. Parametric driving of Josephson plasma oscillations by optically excited phonons was also discussed in Refs.~\cite{Denny2015, Okamoto2016, Okamoto2017, Michael2020}.

\begin{figure}[!t]
	\includegraphics[scale=1]{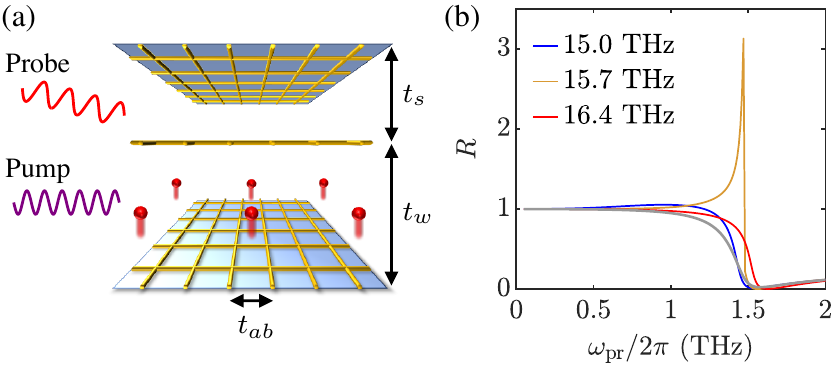}
	\caption{Phonon mediated amplification of terahertz radiation in a bilayer cuprate. (a) Schematic illustration of the pump-probe dynamics in a bilayer cuprate. The superconducting order parameter is discretized on a layered lattice. The pump excites a phonon mode, represented by the red atoms moving along the $c$ axis. Thus, the interlayer tunneling coefficients $t_{s,w}$ become time-dependent, which modifies the plasmonic response to the $c$-axis polarized probe. (b) Reflectivity at normal incidence for different pump frequencies at $T=0$. The modulation amplitudes are $\Lambda_s= 0.2$ and $\Lambda_w= 0.8$. The gray line corresponds to the undriven case. The probe strength is $E_{\mathrm{pr}}= 1~\mathrm{kV/cm}$. The lower and upper Josephson plasma frequencies are $\omega_{\mathrm{J1}}/2\pi= 2~\mathrm{THz}$ and $\omega_{\mathrm{J2}}/2\pi= 14.3~\mathrm{THz}$, respectively. The full parameter set is specified in Table~\ref{tab:parameters}.}
	\label{fig:fig4} 
\end{figure}

Specifically, we consider bilayer cuprates utilizing a particle-hole symmetric $U(1)$ lattice gauge theory in three dimensions \cite{Homann2020, Homann2021}. We formulate a Lagrangian with dynamical and static terms on an anisotropic lattice that corresponds to a bilayer structure as illustrated in Fig.~\ref{fig:fig4}(a).
The static part of the Lagrangian resembles the Ginzburg-Landau free energy \cite{Ginzburg1950}. That is, we describe the Cooper pairs as a condensate of interacting bosons with charge $-2e$, represented by the complex field $\psi_{\mathbf{r}}$. This model is suitable for simulating the coupled dynamics of the order parameter of the superconducting state and the electromagnetic field at temperatures below $T_c$.

The order parameter $\psi_{\mathbf{r}}(t)$ is located on the lattice sites. According to the Peierls substitution, each component of the electromagnetic vector potential $A_{k,\mathbf{r}}(t)$ is defined on the bond between the site $\mathbf{r}$ and its nearest neighbor in the $k \in \{x,y,z\}$ direction. The intra- and interbilayer spacings $d_{s,w}$ are taken as the distances between the CuO$_2$ planes in the crystal, and the in-plane discretization length $d_{ab}$ is introduced as a short-range cutoff of the order of the in-plane coherence length.
The bilayer structure results in the appearance of two Josephson plasma modes. The lower Josephson plasma resonance is dominated by interbilayer currents, due to the interbilayer tunneling energy $t_w$. The upper Josephson plasma resonance, on the other hand, is dominated by intrabilayer currents, due to the intrabilayer tunneling energy $t_s$. We choose the tunneling coefficients $t_s$ and $t_w$ to yield realistic values for the Josephson plasma frequencies. The in-plane tunneling coefficient $t_{ab}$ does not only define an in-plane plasma frequency but also sets the critical temperature of the system. Note that we suppose the $z$~direction to be aligned with the $c$~axis of the crystal.
The Lagrangian of the lattice gauge model is
\begin{equation} \label{eq:Lagrangian}
	\mathcal{L} = \mathcal{L}_{\mathrm{sc}} + \mathcal{L}_{\mathrm{em}} + \mathcal{L}_{\mathrm{kin}} .
\end{equation}
The first term is the $|\psi|^4$ model of the superconducting condensate in the absence of Cooper pair tunneling,
\begin{equation}\label{eq:LSC}
	\mathcal{L}_{\mathrm{sc}} = \sum_{\mathbf{r}} K \hbar^2 | \partial_{t} \psi_{\mathbf{r}} |^2 + \mu | \psi_{\mathbf{r}} |^2 - \frac{g}{2} | \psi_{\mathbf{r}} |^4 ,
\end{equation}
with the fixed Ginzburg-Landau coefficients $\mu$ and $g$. The coefficient $K$ describes the magnitude of the dynamical term \cite{Pekker2015,Tsuji2015}.

The electromagnetic part $\mathcal{L}_{\mathrm{em}}$ is the Lagrangian of the free electromagnetic field on a lattice, modified by the screening due to bound charges in the material,
\begin{equation} \label{eq:HEM}
	\mathcal{L}_{\mathrm{em}} = \sum_{k,\mathbf{r}} \frac{\kappa_{k,\mathbf{r}} \epsilon_{k,\mathbf{r}} \epsilon_0}{2} E_{k,\mathbf{r}}^2 - \frac{\kappa_{z,\mathbf{r}}}{\kappa_{k,\mathbf{r}} \beta_{k,\mathbf{r}}^2 \mu_0} \Bigl[1 - \cos\bigl(\beta_{k,\mathbf{r}} B_{k,\mathbf{r}} \bigr) \Bigr] ,
\end{equation}
where $E_{k,\mathbf{r}}$ denotes the $k$~component of the electric field. Note that we choose the temporal gauge for our calculations, i.e., $E_{k,\mathbf{r}} = -\partial_{t} A_{k,\mathbf{r}}$. The magnetic field components $B_{k,\mathbf{r}}= \epsilon_{klm}\delta_l A_{m,\mathbf{r}}$ are centered on the plaquettes of the lattice. We calculate the spatial derivatives according to $\delta_l A_{m,\mathbf{r}} = (A_{m,\mathbf{r'}(l)}-A_{m,\mathbf{r}})/d_{l,\mathbf{r}}$, where $\mathbf{r'}(l)$ is the neighboring site of $\mathbf{r}$ in the $l$~direction. The discretization lengths are $d_{x,\mathbf{r}}= d_{y,\mathbf{r}}= d_{ab}$ for in-plane junctions, $d_{z,\mathbf{r}}= d_s$ for intrabilayer junctions, and $d_{z,\mathbf{r}}= d_w$ for interbilayer junctions.
The background dielectric constants are $\epsilon_{x,\mathbf{r}}= \epsilon_{y,\mathbf{r}}= \epsilon_{ab}$ for in-plane junctions, $\epsilon_{z,\mathbf{r}}= \epsilon_s$ for intrabilayer junctions, and $\epsilon_{z,\mathbf{r}}= \epsilon_w$ for interbilayer junctions. The other prefactors in Eq.~\eqref{eq:HEM} account for the anisotropic lattice geometry. Introducing $d_c = (d_s + d_w)/2$, we write $\kappa_{x,\mathbf{r}}= \kappa_{y,\mathbf{r}}= 1$ and $\kappa_{z,\mathbf{r}}= d_{z,\mathbf{r}}/d_c$, while $\beta_{x,\mathbf{r}}= \beta_{y,\mathbf{r}}= 2ed_{ab}d_{z,\mathbf{r}}/\hbar$ and $\beta_{z,\mathbf{r}}= 2ed_{ab}^2/\hbar$.

The kinetic part of the Lagrangian is given by
\begin{equation} \label{eq:LEK}
	\mathcal{L}_{\mathrm{kin}} = - \sum_{k,\mathbf{r}} t_{k,\mathbf{r}} |\psi_{\mathbf{r'}(k)} - \psi_{\mathbf{r}} e^{i a_{k,\mathbf{r}}}|^2 .
\end{equation}
The unitless vector potential $a_{k,\mathbf{r}}= -2e d_{k,\mathbf{r}} A_{k,\mathbf{r}}/\hbar$ directly couples to the phase of the order parameter. Thus, it does not only ensure the local gauge-invariance of $\mathcal{L}_{\mathrm{kin}}$, but it also gives rise to a nonlinear coupling between the order parameter and the electromagnetic field. This coupling accounts for the Coulomb interaction between the Cooper pairs.
The Lagrangian \eqref{eq:Lagrangian} is particle-hole symmetric due to its invariance under $\psi_{\mathbf{r}} \rightarrow \psi_{\mathbf{r}}^*$.

We add damping terms and Langevin noise to the equations of motion, which are given by the Euler-Lagrange equations. This enables us to numerically determine the time evolution of the order parameter and the vector potential at zero and nonzero temperature. We employ periodic boundary conditions and integrate the stochastic differential equations using Heun's method with a step size of $\Delta t= 1.6~\mathrm{as}$. To mimic the effect of a driven phonon mode, we make the interlayer tunneling coefficients time-dependent \cite{Okamoto2016, Okamoto2017}, i.e.,
\begin{equation}
t_{s,w} \rightarrow t_{s,w} \left[1 \pm \Lambda_{s,w} \cos(\omega_{\mathrm{dr}}t) \right] .
\end{equation}
This captures a phononic excitation with a wavelength that is large compared to the system size of the simulation. As before, the reflectivity is calculated numerically by adding a probe to the equations of motion for the $z$~component of the electromagnetic vector potential. We assume the existence of a suitable phonon resonance such that $\omega_{\mathrm{dr}}$ is blue-detuned from the upper Josephson plasma frequency $\omega_{\mathrm{J2}}$.

\begin{figure*}[!t]
	\includegraphics[scale=1]{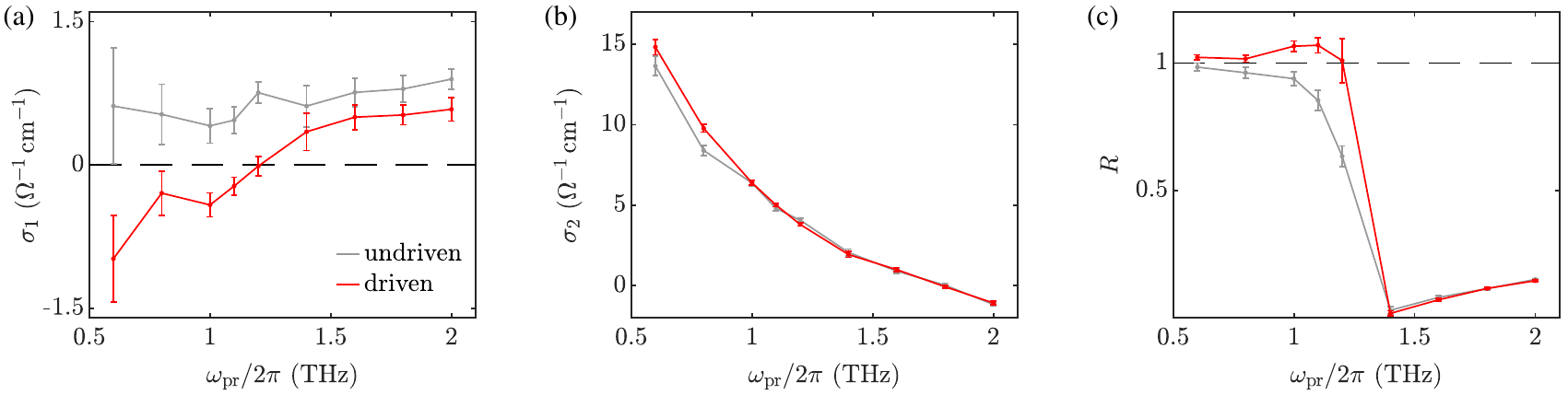}
	\caption{Phonon mediated amplification of terahertz radiation in a bilayer cuprate at $T= 5~\mathrm{K} \sim 0.2 T_c$. (a) Real part of the optical conductivity. (b) Imaginary part of the optical conductivity. (c) Reflectivity at normal incidence. The pump frequency is $\omega_{\mathrm{dr}}/2\pi= 14.8~\mathrm{THz}$, and the modulation amplitudes are $\Lambda_s= 0.2$ and $\Lambda_w= 0.8$. The error bars indicate the standard errors of the ensemble averages. The probe strength is $E_{\mathrm{pr}}= 30~\mathrm{kV/cm}$. The bilayer system is the same as in Fig.~\ref{fig:fig4}. The upper Josephson plasma frequency is shifted to $\omega_{\mathrm{J2}} \approx 13.8~\mathrm{THz}$ due to thermal fluctuations.}
	\label{fig:fig5} 
\end{figure*}

In Fig.~\ref{fig:fig4}(b), we show that the phonon mediated pump has a similar effect as the plasmonic excitations discussed before. This analogy derives from tunneling terms of the form $\sim$$t_{k,\mathbf{r}} A_{k,\mathbf{r}}^2$ in the Lagrangian. That is, the parametric amplification of the terahertz probe is enabled by the cubic coupling of the tunneling coefficients and the vector potential. In contrast to the case of Higgs mode mediated amplification, the pump does not primarily couple to the vector potential but to the tunneling coefficients, which models the excited phonon mode.
Here, the maximum gain is realized when the detuning $\omega_{\mathrm{dr}}-\omega_{\mathrm{J2}}$ approaches the frequency of the lower reflectivity edge at $\omega_{\mathrm{J1}}/\sqrt{2}$. The amplification of terahertz radiation is feasible up to probe strengths of $\sim$$100~\mathrm{kV/cm}$; see Appendix~\ref{sec:strengths}. Note that the amplification mechanism also works if the frequency of the excited phonon mode is slightly blue-detuned from the lower Josephson plasma frequency. However, this requires a pump frequency of the order of 1~THz \cite{Basov2005}, which is in the frequency range that lacks suitable radiation sources.

Finally, we investigate the phonon mediated amplification of terahertz signals at nonzero temperature $T$ by simulating an ensemble of several hundred trajectories for a bilayer system of $40 \times 40 \times 4$ sites. To obtain the $c$-axis conductivity $\sigma(\omega_{\mathrm{pr}})$ for a single trajectory, we evaluate the sample averages of the $c$-axis current $J_z(\omega_{\mathrm{pr}})$ and the $c$-axis electric field $E_z(\omega_{\mathrm{pr}})$. We then take the ensemble average of $\sigma(\omega_{\mathrm{pr}})$ and calculate the reflectivity $R(\omega_{\mathrm{pr}})$ using Eq.~\eqref{eq:reflectivity}. As shown in Figs.~\ref{fig:fig5}(a) and \ref{fig:fig5}(c), the phonon mediated amplification of terahertz radiation is effective at $\sim$20\% of $T_c$ despite the thermal broadening of the parametric resonance expected at $\omega_{\mathrm{pr}}/2\pi \approx 1~\mathrm{THz}$. Importantly, the real part of the conductivity is negative at frequencies around 1 THz and below, leading to a reflectivity $R>1$ in this regime. Additionally, we observe a parametric enhancement of the imaginary part of the low-frequency conductivity in Fig.~\ref{fig:fig5}(b); see also Refs.~\cite{Denny2015, Okamoto2016, Okamoto2017, Homann2021}.

\section{Discussion and outlook}
In conclusion, we propose a terahertz amplification technology based on parametric amplification in high-$T_c$ superconductors, utilizing an optical pump mechanism. A key feature of the amplifier and its underlying mechanism is that the enhancement of the reflectivity is controlled via the pump frequency and the pump strength.
Superconductors are promising candidates to induce a reflectivity $R>1$ because their low-frequency reflectivity is close to 1 in equilibrium. We emphasize, however, that the mechanism we put forth can not only be realized in cuprate superconductors but also in other materials with collective modes that couple nonlinearly to light. The parametric amplification of terahertz signals is limited by the finite penetration depth of the pump, which is smaller than the penetration depth of the probe in many cases \cite{Buzzi2021, Hu2014, Kaiser2014}. To reduce the mismatch of the penetration depths of the pump and the probe, we propose to choose a large incident angle for the probe beam while orienting the pump beam parallel to the surface normal; see Fig.~\ref{fig:fig1}. As mentioned in the introduction, we recommend to implement an amplifier in pulsed operation once net optical gain from a light-driven solid is achieved. This would also be advantageous with regards to heating effects, which are further discussed in Appendix~\ref{sec:heating}.

Our proposed terahertz amplifier advances pump-probe experiments on high-$T_c$ superconductors towards a potential application. It motivates a demonstration of a stable and sufficiently strong enhancement of the reflectivity above 1 and the design of an optical cavity as shown in Fig.~\ref{fig:fig1}, with the purpose of developing coherent radiation sources in the terahertz regime.

\begin{acknowledgments}

We thank Reinhold Kleiner, Lukas Broers, and Jim Skulte for stimulating discussions.
This work is supported by the Deutsche Forschungsgemeinschaft (DFG) in the framework of SFB~925, Project No.~170620586, and the Cluster of Excellence ``Advanced Imaging of Matter" (EXC~2056), Project No.~390715994.

\end{acknowledgments}

\appendix

\section{Analytical estimate of Higgs mode mediated amplification in monolayer cuprates}
\label{sec:analytical}
\begin{figure}[!b]
	\centering
	\includegraphics[scale=1]{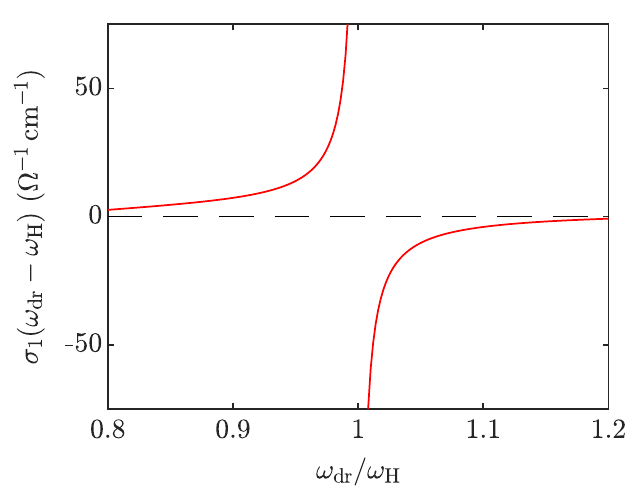}
	\caption{Analytical estimate of the optical conductivity of a monolayer cuprate, in which Josephson plasma oscillations are optically driven. The negative conductivity for $\omega_{\mathrm{dr}} \gtrsim \omega_{\mathrm{H}}$ indicates the Higgs mode mediated amplification of terahertz radiation. As in Fig.~\ref{fig:fig2}, the Josephson plasma frequency is $\omega_{\mathrm{J}}/2\pi= 2~\mathrm{THz}$ and the Higgs frequency is $\omega_{\mathrm{H}}/2\pi= 6~\mathrm{THz}$. The remaining parameters are $E_0= 300~\mathrm{kV/cm}$,  $\gamma_{\mathrm{J}}/2\pi= 0.25~\mathrm{THz}$, $\gamma_{\mathrm{H}}/2\pi= 1~\mathrm{THz}$, $\alpha=1$, $\epsilon_{\infty}=4$, and $d= 10~\text{\AA}$.}
	\label{fig:analytical} 
\end{figure}

Neglecting all nonlinear terms except for the quadratic coupling between the Higgs mode $h$ and the plasma mode $\theta$ in Eqs.~\eqref{eq:eom1} and \eqref{eq:eom2}, we find
\begin{align}
	\ddot{\theta} + \gamma_{\mathrm{J}} \dot{\theta} + \omega_{\mathrm{J}}^2 \theta + 2 \omega_{\mathrm{J}}^2 \theta h &= j , \label{eq:eom1_quad} \\
	\ddot{h} + \gamma_{\mathrm{H}} \dot{h} + \omega_{\mathrm{H}}^2 h + \alpha \omega_{\mathrm{J}}^2 \theta^2 &= 0 , \label{eq:eom2_quad}
\end{align}
as in Refs.~\cite{Homann2020, Homann2021}. Now, we expand $j$, $\theta$, and $h$ in the form
\begin{equation}
	f= f^{(0)} + \lambda f^{(1)} + \lambda^2 f^{(2)} + \lambda^3 f^{(3)} + \mathcal{O}(\lambda^4) ,
\end{equation}
where $\lambda \ll 1$ is a small expansion parameter. We take the current $j$ induced by the pump and the probe as
\begin{equation}
	j^{(1)}= j_{\mathrm{dr},1} e^{- i \omega_{\mathrm{dr}} t} + j_{\mathrm{pr},1} e^{- i \omega_{\mathrm{pr}} t} + \mathrm{c.c.} .
\end{equation}
Hence, there are no zeroth order contributions and we obtain
\begin{align}
	\theta^{(1)} &= \theta_{\mathrm{dr},1} e^{- i \omega_{\mathrm{dr}} t} + \theta_{\mathrm{pr},1} e^{- i \omega_{\mathrm{pr}} t} + \mathrm{c.c.} , \\
	h^{(1)} &= 0
\end{align}
in first order, where
\begin{align}
	\theta_{\mathrm{dr,1}} &= \frac{j_{\mathrm{dr},1}}{\omega_{\mathrm{J}}^2 - \omega_{\mathrm{dr}}^2 - i \gamma_{\mathrm{J}} \omega_{\mathrm{dr}}} , \\
	\theta_{\mathrm{pr,1}} &= \frac{j_{\mathrm{pr},1}}{\omega_{\mathrm{J}}^2 - \omega_{\mathrm{pr}}^2 - i \gamma_{\mathrm{J}} \omega_{\mathrm{pr}}} . \label{eq:den1}
\end{align}
\begin{widetext}
In second order, we have
\begin{align}
	\theta^{(2)} &= 0 , \\
	h^{(2)} &= h_0 + h_1 e^{- 2 i \omega_{\mathrm{dr}} t} + h_2 e^{- 2 i \omega_{\mathrm{pr}} t} + h_3 e^{- i(\omega_{\mathrm{dr}} - \omega_{\mathrm{pr}}) t} + h_4 e^{- i(\omega_{\mathrm{dr}} + \omega_{\mathrm{pr}}) t} + \mathrm{c.c.} ,
\end{align}
where
\begin{align}
	h_0 &= - \frac{2 \alpha \omega_{\mathrm{J}}^2}{\omega_{\mathrm{H}}^2} \big( |\theta_{\mathrm{dr},1}|^2 + |\theta_{\mathrm{pr},1}|^2 \big) , \\
	h_1 &= \frac{\alpha \omega_{\mathrm{J}}^2 \theta_{\mathrm{dr},1}^2}{4 \omega_{\mathrm{dr}}^2 - \omega_{\mathrm{H}}^2 + 2 i \gamma_{\mathrm{H}} \omega_{\mathrm{dr}}} , \\
	h_2 &= \frac{\alpha \omega_{\mathrm{J}}^2 \theta_{\mathrm{pr},1}^2}{4 \omega_{\mathrm{pr}}^2 - \omega_{\mathrm{H}}^2 + 2 i \gamma_{\mathrm{H}} \omega_{\mathrm{pr}}} , \\
	h_3 &= \frac{2 \alpha \omega_{\mathrm{J}}^2 \theta_{\mathrm{dr},1} \theta_{\mathrm{pr},1}^*}{(\omega_{\mathrm{dr}} - \omega_{\mathrm{pr}})^2 - \omega_{\mathrm{H}}^2 + i \gamma_{\mathrm{H}} (\omega_{\mathrm{dr}} - \omega_{\mathrm{pr}})} , \label{eq:den2} \\
	h_4 &= \frac{2 \alpha \omega_{\mathrm{J}}^2 \theta_{\mathrm{dr},1} \theta_{\mathrm{pr},1}}{(\omega_{\mathrm{dr}} + \omega_{\mathrm{pr}})^2 - \omega_{\mathrm{H}}^2 + i \gamma_{\mathrm{H}} (\omega_{\mathrm{dr}} + \omega_{\mathrm{pr}})} . \label{eq:den3}
\end{align}
In third order, we find the following correction for the vector potential at the probe frequency,
\begin{equation} \label{eq:den4}
	\theta_{\mathrm{pr},3}=  \frac{2 \omega_{\mathrm{J}}^2 \big( h_0 \theta_{\mathrm{pr},1} + h_2 \theta_{\mathrm{pr},1}^* + h_3^* \theta_{\mathrm{dr},1} + h_4 \theta_{\mathrm{dr},1}^* \big)}{\omega_{\mathrm{pr}}^2 - \omega_{\mathrm{J}}^2 + i \gamma_{\mathrm{J}} \omega_{\mathrm{pr}}} .
\end{equation}
We consider a probe with $|j_{\mathrm{pr},1}| \ll |j_{\mathrm{dr},1}|$ and $\omega_{\mathrm{pr}}= \omega_{\mathrm{dr}} - \omega_{\mathrm{H}}$ such that we can neglect the $h_2$ term. Moreover, we assume near-resonant driving, i.e., $\omega_{\mathrm{dr}} \simeq \omega_{\mathrm{H}}$, to simplify the denominators in Eqs.~\eqref{eq:den2} and \eqref{eq:den3},
\begin{align}
	\begin{split}
		\theta_{\mathrm{pr},3} &\approx  \frac{ -4 \alpha \omega_{\mathrm{J}}^4 |\theta_{\mathrm{dr},1}|^2 \theta_{\mathrm{pr},1}}{\omega_{\mathrm{pr}}^2 - \omega_{\mathrm{J}}^2 + i \gamma_{\mathrm{J}} \omega_{\mathrm{pr}}} \Biggl( \frac{1}{\omega_{\mathrm{H}}^2} + \frac{2}{i \gamma_{\mathrm{H}} \omega_{\mathrm{H}}} \Biggr) \\
		&\approx \frac{ 4 \alpha \omega_{\mathrm{J}}^4 |j_{\mathrm{dr},1}|^2 j_{\mathrm{pr},1} (\gamma_{\mathrm{H}} - 2i \omega_{\mathrm{H}}) }{ \gamma_{\mathrm{H}} \omega_{\mathrm{H}}^2 \bigl[ (\omega_{\mathrm{dr}} - \omega_{\mathrm{H}})^2 - \omega_{\mathrm{J}}^2 + i \gamma_{\mathrm{J}} (\omega_{\mathrm{dr}} - \omega_{\mathrm{H}}) \bigr]^2 \bigl[ (\omega_{\mathrm{dr}}^2 - \omega_{\mathrm{J}}^2)^2 + \gamma_{\mathrm{J}}^2 \omega_{\mathrm{dr}}^2 \bigr] } .
	\end{split}
\end{align}
As $\dot{\theta}= 2edE/\hbar$ \cite{Josephson1962}, the optical conductivity is given by
\begin{equation}
	\sigma(\omega_{\mathrm{pr}}) 
	= \frac{i \epsilon_{\infty} \epsilon_0 j (\omega_{\mathrm{pr}})}{\omega_{\mathrm{pr}} \theta(\omega_{\mathrm{pr}})} 
	= \frac{i \epsilon_{\infty} \epsilon_0 \lambda j_{\mathrm{pr},1}}{\omega_{\mathrm{pr}} (\lambda \theta_{\mathrm{pr},1} + \lambda^3 \theta_{\mathrm{pr},3})} ,
\end{equation}
and we obtain
\begin{align}
	\begin{split} \label{eq:sigma}
		&\sigma(\omega_{\mathrm{pr}}= \omega_{\mathrm{dr}} - \omega_{\mathrm{H}}) \\
		&\approx \frac{ -i \epsilon_{\infty} \epsilon_0 \gamma_{\mathrm{H}} \omega_{\mathrm{H}}^2 \bigl[ (\omega_{\mathrm{dr}} - \omega_{\mathrm{H}})^2 - \omega_{\mathrm{J}}^2 + i \gamma_{\mathrm{J}} (\omega_{\mathrm{dr}} - \omega_{\mathrm{H}}) \bigr]^2 \bigl[ (\omega_{\mathrm{dr}}^2 - \omega_{\mathrm{J}}^2)^2 + \gamma_{\mathrm{J}}^2 \omega_{\mathrm{dr}}^2 \bigr] (\omega_{\mathrm{dr}} - \omega_{\mathrm{H}})^{-1}}{ \gamma_{\mathrm{H}} \omega_{\mathrm{H}}^2 \bigl[ (\omega_{\mathrm{dr}} - \omega_{\mathrm{H}})^2 - \omega_{\mathrm{J}}^2 + i \gamma_{\mathrm{J}} (\omega_{\mathrm{dr}} - \omega_{\mathrm{H}}) \bigr] \bigl[ (\omega_{\mathrm{dr}}^2 - \omega_{\mathrm{J}}^2) + \gamma_{\mathrm{J}}^2 \omega_{\mathrm{dr}}^2 \bigr] - 4 \alpha \omega_{\mathrm{J}}^4 |j_{\mathrm{dr}}|^2 (\gamma_{\mathrm{H}} - 2i \omega_{\mathrm{H}}) } ,
	\end{split}
\end{align}
with the original pump amplitude $j_{\mathrm{dr}}= \lambda j_{\mathrm{dr},1}$. Taking $j_{\mathrm{dr}}=0$ leads to the equilibrium solution
\begin{equation}
	\sigma_1 (\omega_{\mathrm{pr}}= \omega_{\mathrm{dr}} - \omega_{\mathrm{H}}) = \epsilon_{\infty} \epsilon_0 \gamma_{\mathrm{J}} .
\end{equation}
\end{widetext}
for the real part of the conductivity. In the above calculation, $\omega_{\mathrm{pr}}$ is formally negative for $\omega_{\mathrm{dr}} < \omega_{\mathrm{H}}$. However, our analytical prediction has the property that $\sigma_1(\omega_{\mathrm{pr}})=\sigma_1(-\omega_{\mathrm{pr}})$, as characteristic for Fourier transforms of real quantities.
In Fig.~\ref{fig:analytical}, the real part of the conductivity at $\omega_{\mathrm{pr}}= \omega_{\mathrm{dr}} - \omega_{\mathrm{H}}$ is displayed as a function of the pump frequency according to our analytical estimate in Eq.~\eqref{eq:sigma}. The field strength  $E_0$ of the applied electric field gives rise to the pump amplitude $|j_{\mathrm{dr}}|= ed \omega_{\mathrm{dr}} E_0 / \hbar \epsilon_{\infty}$. Consistent with the numerical results, we find a negative conductivity when the pump frequency is slightly blue-detuned from the Higgs frequency. On the other hand, the conductivity is positive for red-detuned pump frequencies. A quantitative comparison to the numerical results reveals notable deviations, which are due to the approximations made in the derivation of Eq.~\eqref{eq:sigma}.

\section{Equations of motion for the three-dimensional lattice gauge model}
Including damping terms and thermal fluctuations, the equations of motion read
\begin{align}
	\partial_t^2 \psi_{\mathbf{r}} &= \frac{1}{K \hbar^2} \frac{\partial \mathcal{L}}{\partial \psi_{\mathbf{r}}^*} - \gamma_{\mathrm{H}} \partial_t \psi_{\mathbf{r}} + \xi_{\mathbf{r}} , \\
	\partial_t^2 A_{x, \mathbf{r}} &= \frac{1}{\epsilon_{ab} \epsilon_0} \frac{\partial \mathcal{L}}{\partial A_{x, \mathbf{r}}} - \gamma_{ab} \partial_t A_{x, \mathbf{r}} + \eta_{x, \mathbf{r}} , \\
	\partial_t^2 A_{y, \mathbf{r}} &= \frac{1}{\epsilon_{ab} \epsilon_0} \frac{\partial \mathcal{L}}{\partial A_{y, \mathbf{r}}} - \gamma_{ab} \partial_t A_{y, \mathbf{r}} + \eta_{y, \mathbf{r}} , \\
	\partial_t^2 A_{z, \mathbf{r}} &= \frac{1}{\kappa_{z,\mathbf{r}} \epsilon_{z,\mathbf{r}} \epsilon_0} \frac{\partial \mathcal{L}}{\partial A_{z, \mathbf{r}}} - \gamma_{z,\mathbf{r}} \partial_t A_{z, \mathbf{r}} + \eta_{z, \mathbf{r}} ,
\end{align}
where $\xi_{\mathbf{r}}$ and $\boldsymbol{\eta}_{\mathbf{r}}$ represent the thermal fluctuations of the superconducting order parameter and the vector potential, respectively. These Langevin noise terms have a white Gaussian distribution with zero mean. The damping coefficients of the intra- and interbilayer electric fields are $\gamma_s$ and $\gamma_w$, respectively. To satisfy the fluctuation-dissipation theorem, we take the noise of the order parameter as
\begin{align}
	\langle \mathrm{Re} \{{\xi_{\mathbf{r}} (t)}\} \mathrm{Re} \{{\xi_{\mathbf{r'}} (t')}\} \rangle &= \frac{\gamma_{\mathrm{H}} k_{\mathrm{B}} T}{K \hbar^2 V_0} \delta_{\mathbf{r}\mathbf{r'}} \delta(t-t') \, , \\
	\langle \mathrm{Im} \{{\xi_{\mathbf{r}} (t)}\} \mathrm{Im} \{{\xi_{\mathbf{r'}} (t')}\} \rangle &= \frac{\gamma_{\mathrm{H}} k_{\mathrm{B}} T}{K \hbar^2 V_0} \delta_{\mathbf{r}\mathbf{r'}} \delta(t-t') \, , \\
	\langle \mathrm{Re} \{{\xi_{\mathbf{r}} (t)}\} \mathrm{Im} \{{\xi_{\mathbf{r'}} (t')}\} \rangle &= 0 ,
\end{align}
where $V_0= d_{ab}^2 d_c$. The noise correlations for the vector potential are
\begin{align}
	\langle \eta_{x,\mathbf{r}} (t) \eta_{x,\mathbf{r'}} (t') \rangle &= \frac{2 \gamma_{ab} k_{\mathrm{B}} T}{\epsilon_{ab} \epsilon_0 V_0} \delta_{\mathbf{r}\mathbf{r'}} \delta(t-t') , \\
	\langle \eta_{y,\mathbf{r}} (t) \eta_{y,\mathbf{r'}} (t') \rangle &= \frac{2 \gamma_{ab} k_{\mathrm{B}} T}{\epsilon_{ab} \epsilon_0 V_0} \delta_{\mathbf{r}\mathbf{r'}} \delta(t-t') , \\
	\langle \eta_{z,\mathbf{r}} (t) \eta_{z,\mathbf{r'}} (t') \rangle &= \frac{2 \gamma_{z,\mathbf{r}} k_{\mathrm{B}} T}{\kappa_{z,\mathbf{r}} \epsilon_{z,\mathbf{r}} \epsilon_0 V_0} \delta_{\mathbf{r}\mathbf{r'}} \delta(t-t') .
\end{align}

\section{Simulation parameters of the bilayer cuprate}
\label{sec:parameters}
In this work, we simulate a bilayer cuprate with $40 \times 40 \times 4$ sites, choosing the parameters summarized in Table~\ref{tab:parameters}. Our choice of $\mu$ and $g$ implies an equilibrium condensate density $n_0= \mu/g = 2 \times 10^{21}~\mathrm{cm^{-3}}$ at $T=0$. The bilayer system has two longitudinal $c$-axis plasma modes. Their eigenfrequencies are
\begin{widetext}
	\begin{equation}
		\omega_{\mathrm{J1,J2}}^2 = \biggl( \frac{1}{2}+ \alpha_{s} \biggr) \Omega_{s}^2 + \biggl( \frac{1}{2}+ \alpha_{w} \biggr) \Omega_{w}^2 \mp \sqrt{ \biggl[\biggl( \frac{1}{2}+ \alpha_{s} \biggr) \Omega_{s}^2 - \biggl( \frac{1}{2}+ \alpha_{w} \biggr) \Omega_{w}^2 \biggr]^2 + 4 \alpha_{s} \alpha_{w} \Omega_{s}^2 \Omega_{w}^2 } ,
	\end{equation}
\end{widetext}
as follows from a sine-Gordon analysis \cite{Marel2001, Koyama2002}. Here we introduced the bare plasma frequencies of the strong and weak junctions
\begin{equation}
	\Omega_{s,w}= \sqrt{\frac{8 t_{s,w} n_0  e^2 d_c d_{s,w}}{\epsilon_{s,w} \epsilon_0 \hbar^2}} ,
\end{equation}
where $d_c= (d_s + d_w)/2$. The capacitive coupling constants are given by
\begin{equation}
	\alpha_{s,w}= \frac{\epsilon_{s,w} \epsilon_0}{8 K n_0 e^2 d_c d_{s,w}} .
\end{equation}
Besides, there is a transverse $c$-axis plasma mode with the eigenfrequency
\begin{equation}
	\omega_{\mathrm{T}}^2 = \frac{1+ 2\alpha_s + 2\alpha_w}{\alpha_s + \alpha_w} \, \Bigl( \alpha_{s} \Omega_s^2 + \alpha_w \Omega_w^2 \Bigr) .
\end{equation}
We have $\alpha_s \approx 1.5$, $\alpha_w \approx 3.0$, $\omega_{\mathrm{J1}}/2\pi \approx 2.0~\mathrm{THz}$, $\omega_{\mathrm{J2}}/2\pi \approx 14.3~\mathrm{THz}$, and $\omega_{\mathrm{T}}/2\pi \approx 13.2~\mathrm{THz}$ for the parameters specified in Table~\ref{tab:parameters}.
\begin{table}[!b]
	\caption{Model parameters of the simulated bilayer cuprate.}
	\renewcommand{\arraystretch}{1.5}
	\begin{tabular}{lr}
		\hline
		$K~(\text{meV}^{-1})$ & $1.9 \times 10^{-5}$ \\
		$\mu~(\text{meV})$ & $6.0 \times 10^{-3}$ \\
		$g~(\text{meV} \, \text{\AA}^3)$ & 3.0 \\
		\hline
		$\gamma_{\mathrm{H}}/2\pi~(\mathrm{THz})$ & 1.0 \\
		$\gamma_{ab}/2\pi~(\mathrm{THz})$ & 7.0 \\
		$\gamma_s/2\pi~(\mathrm{THz})$ & 1.5 \\
		$\gamma_w/2\pi~(\mathrm{THz})$ & 0.25 \\
		\hline
		$\epsilon_{ab}$ & 4 \\
		$\epsilon_s$ & 2 \\
		$\epsilon_w$ & 8 \\
		$d_{ab}~(\text{\AA})$ & 15 \\
		$d_s~(\text{\AA})$ & 4 \\
		$d_w~(\text{\AA})$ & 8 \\
		$t_{ab}~(\text{meV})$ & $5.2 \times 10^{-1}$ \\
		$t_s~(\text{meV})$ & $\qquad 2.4 \times 10^{-2}$ \\
		$t_w~(\text{meV})$ & $\qquad 1.7 \times 10^{-3}$ \\
		\hline
	\end{tabular}
	\renewcommand{\arraystretch}{1}
	\label{tab:parameters}
\end{table}
The in-plane plasma frequency is
\begin{equation}
	\omega_{ab}= \sqrt{\frac{8t_{ab} n_0 e^2 d_{ab}^2}{\epsilon_{ab} \epsilon_0 \hbar^2}} \approx 2\pi \times 70~\mathrm{THz} ,
\end{equation}
and the Higgs frequency is
\begin{equation}
	\omega_{\mathrm{H}}= \sqrt{\frac{2\mu}{K \hbar^2}} \approx 2\pi \times 6.1~\mathrm{THz} .
\end{equation}
The average background dielectric constant along the $c$~axis is
\begin{equation}
	\epsilon_{\infty}= \frac{(d_s + d_w) \epsilon_s \epsilon_w}{d_s \epsilon_w + d_w \epsilon_s} = 4 .
\end{equation}

\begin{figure}[!b]
	\centering
	\includegraphics[scale=1]{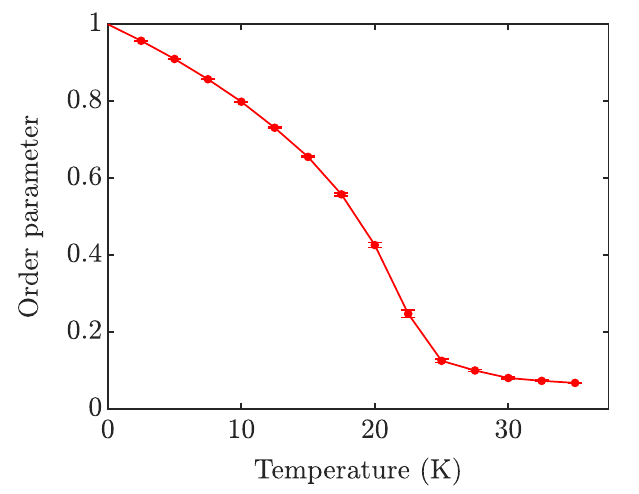}
	\caption{Phase transition of a bilayer cuprate with $40 \times 40 \times 4$ sites and the parameters specified in Table~\ref{tab:parameters}. The error bars indicate the standard errors of the ensemble averages.}
	\label{fig:finiteTemp} 
\end{figure}

\begin{figure*}[!t]
	\centering
	\includegraphics[scale=1]{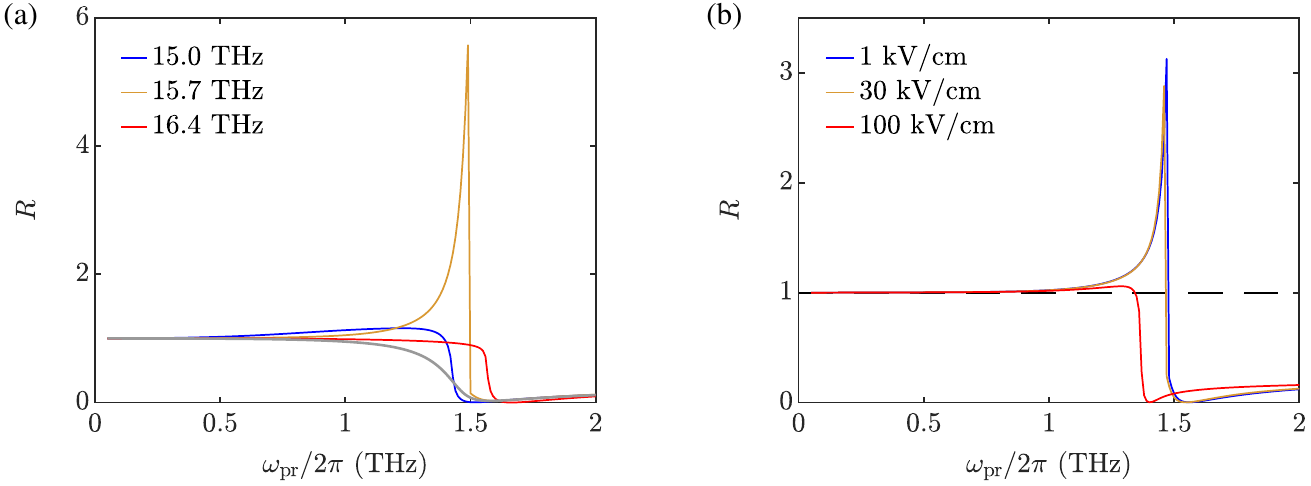}
	\caption{Reflectivity of a bilayer cuprate in the presence of a phonon mediated pump at $T=0$. (a) Reflectivity at normal incidence for different pump frequencies. The modulation amplitudes of the interlayer tunneling coefficients are $\Lambda_s=0.3$ and $\Lambda_w=1$. The gray line corresponds to the undriven case. (b) Reflectivity at normal incidence for different probe strengths. The pump frequency is $\omega_{\mathrm{dr}}/2\pi= 15.7~\mathrm{THz}$, and the modulation amplitudes are $\Lambda_s= 0.2$ and $\Lambda_w= 0.8$. See Appendix~\ref{sec:parameters} for model parameters.}
	\label{fig:phononAmp} 
\end{figure*}

\section{Thermal phase transition}
The thermal equilibrium at a given temperature is established as follows. We initialize the system in its ground state at $T=0$ and let the dynamics evolve without external driving, influenced only by thermal fluctuations and dissipation. To characterize the phase transition, we introduce the order parameter
\begin{equation}
	O= \frac{ 2 \Big|\sum_{ \{\mathbf{r}, \mathbf{r'}\} \in \mathrm{inter}} \psi_{\mathbf{r'}}^* \psi_{\mathbf{r}} \, \mathrm{e}^{\mathrm{i} a_{z,\mathbf{r}}} \Big| }{\sum_{\mathbf{r}} |\psi_{\mathbf{r}}|^2}.
\end{equation}
The sum in the numerator is taken over all interbilayer bonds, with site $\mathbf{r}$ in the lower layer. Thus, the order parameter measures the gauge-invariant phase coherence of the condensate across different bilayers. In our simulations, this quantity converges to a constant after 10~ps of free time evolution, indicating the onset of thermal equilibrium. For each trajectory, the order parameter is evaluated from the average of 200 measurements within a time interval of 2~ps. Finally, we take the ensemble average of 100 trajectories.
As depicted in Fig.~\ref{fig:finiteTemp}, the temperature dependence of the order parameter is reminiscent of a second order phase transition. Due to the finite size of the simulated system, the order parameter converges to a plateau with nonzero value for high temperatures. At $T_c \sim 25~{\mathrm{K}}$, there is a distinct crossover.

\section{Dependence of phonon mediated amplification on the pump and probe strengths}
\label{sec:strengths}
Figure~\ref{fig:phononAmp}(a) displays the reflectivity of a bilayer cuprate, corresponding to the examples of phonon mediated amplification of terahertz radiation in Fig.~\ref{fig:fig4}(b). While the pump frequencies are the same, the modulation amplitudes of the interlayer tunneling coefficients are larger here.

Next, we investigate the dependence of the reflectivity on the probe strength. As one can see in Fig.~\ref{fig:phononAmp}(b), the results for $E_{\mathrm{pr}}= 1~\mathrm{kV/cm}$ and $E_{\mathrm{pr}}= 30~\mathrm{kV/cm}$ are in very good agreement. There is only a small deviation close to the maximum. This demonstrates that these probe strengths correspond to the linear response regime. For higher probe strenths, however, the amplification peak in the reflectivity decreases and the plasma edge is shifted to lower frequencies. Remarkably, the reflectivity still exceeds 1 for probe frequencies slightly above 1 THz when the probe strength reaches 100 kV/cm.

\section{Sample heating}
\label{sec:heating}
Here, we estimate an upper limit for the heating of the sample by a pump pulse. We consider a cubic YBCO sample with a volume of 1 mm$^3$, corresponding to $N\approx 10^{-5}~\mathrm{mol}$. The specific heat capacity of YBCO at 20 K is $C \approx 4~\mathrm{J \, mol^{-1} K^{-1}}$ \cite{Loram1993}. If an entire face of the sample is irradiated by a laser with a fluence of $u= 20~\mathrm{mJ \, cm^{-2}}$ \cite{Budden2021}, the sample can absorb energy up to an amount of $U= 0.2~\mathrm{mJ}$. Assuming that all the energy is dissipated and converted into heat, we find a temperature increase of
\begin{equation}
	\Delta T= \frac{U}{C N} \approx 5~\mathrm{K} .
\end{equation}
For a material with $T_c \sim 90~\mathrm{K}$, this would be a moderate effect. Due to the finite penetration depth of the pump, the surface of the sample generally heats up disproportionately. Experimental observations on various cuprates indicate robustness against surface heating for pump pulses with a field strength of $\sim$1~MV/cm and a duration of a few hundred femtoseconds \cite{Hu2014, Kaiser2014, Cremin2019}.

\bibliography{biblio}

\end{document}